\documentstyle[10pt,emulateapj]{article}

\def\solm{M$_{\odot}\,$}

\begin{document}     
\vspace{-7cm}
\title{Observing the Formation of the Hubble Sequence in the Great 
Observatories Origins Deep Survey$^{1,2}$}

\author{Christopher J. Conselice$^{3,4}$, Norman A. Grogin$^{5}$, Shardha Jogee$^{6}$, Ray A. Lucas$^{6}$,  Tomas Dahlen$^{6}$,  Duilia de Mello$^{7}$, Jonathan P. Gardner$^{7}$, Bahram Mobasher$^{6}$, Swara Ravindranath$^{6}$}
\altaffiltext{1}{Based on observations taken with the NASA/ESA Hubble
Space Telescope, which is operated by the Association of Universities
for Research in Astronomy, Inc.\ (AURA) under NASA contract NAS5--26555}
\altaffiltext{2}{Based on observations collected at the European Southern 
Observatory, Chile}
\altaffiltext{3}{California Institute of Technology, Mail Code 105-24, Pasadena
CA 91125}
\altaffiltext{4}{NSF Astronomy \& Astrophysics Postdoctoral Fellow}
\altaffiltext{5}{Johns Hopkins University, Baltimore MD}
\altaffiltext{6}{Space Telescope Science Institute, Baltimore MD}
\altaffiltext{7}{Laboratory for Astronomy and Solar Physics, 
Code 681, Goddard Space Flight Center, Greenbelt MD 20771}
\begin{abstract}

Understanding the physical formation of the Hubble sequence remains one of
the most important unsolved astrophysical
problems. Searches for proto-disks and proto-ellipticals can now be 
effectively done using deep
wide-field Hubble Space Telescope images taken with the new
Advanced Camera for Surveys.  Through an analysis of the concentrations ($C$),
asymmetries ($A$) and clumpiness values ($S$) (CAS) of galaxies found in the 
GOODS Field South, we are able to identify objects possibly forming onto
the Hubble sequence.  Using this approach, we
detect a sizeable population of star forming luminous diffuse
objects and star forming luminous asymmetric objects between redshifts 
$0.5 < z < 2$.  These galaxies have extremely low light
concentrations, or high asymmetries, with
absolute magnitudes M$_{\rm B} < -19$.  The luminous diffuse objects
are found in abundance between $z = 1 - 2$, with fewer objects
at $z > 2$ and $z < 1$.  The luminous asymmetric objects
are found at a similar abundance, with a peak
at $z \sim 1$.   We argue that these
galaxies are a subset of modern disks and ellipticals in formation.  
The co-moving volume density of the luminous diffuse objects between 
$z = 1 - 2$ is similar to the local density of bright disk galaxies, 
with values $\sim$ 5 $\times 10^{5}$ Gpc$^{-3}$.  The SEDs of these objects 
are mostly consistent with 
starbursts, or star-forming normal galaxies, with average uncorrected for
extinction star formation rates of $\sim 4$ \solm yr$^{-1}$. These galaxies 
also host 35-40 \% of the star formation activity at $1<z<2$.
We briefly discuss the implications of
these objects for understanding the origin of the Hubble sequence.

\end{abstract}

\section{Introduction}

Understanding the evolution of galaxies is a major unsolved astrophysical
problem, typically divided into two separate well studied regimes.  
One is in the nearby Universe where disks
and ellipticals have been studied in detail for nearly a century.  
Perhaps surprisingly, the other well studied epoch
is at high redshift, $z > 2.5$, where galaxies are selected through 
observational techniques, such
as the Lyman-break criteria (Steidel et al. 1996).  To address
the entire problem of galaxy evolution one needs to understand how high
redshift galaxy populations evolve into the galaxies we see at low redshift. 
That is, we want to know when and how the morphologically 
identifiable  $z \sim 0$ galaxy population first formed. 

The standard picture of galaxy evolution consists of the following.  In
the very early Universe, gas collapsed in dark matter halos, which later cooled
to form stars, creating the first galaxies. Later these dark halos merged
to form larger dark halos, and thus more massive galaxies (e.g., 
Cole et al. 2000; Somerville, Primack \& Faber 2001). In this hierarchical
picture, disks formed around previously existing spheroids through
accretion of gas from the intergalactic medium, while the spheroids
themselves formed through major mergers (e.g., Steinmetz \& Navarro 2002).   
Based on the morphological appearances of galaxies, this merger
epoch ends at about $z \sim 2$ for the brightest and most massive systems
(Conselice et al. 2003).    Yet, we still have very little knowledge of when
or how the modern Hubble sequence came into place.

Galaxies at $z > 2.5$ appear very irregular, with no obvious morphological 
properties similar to ellipticals or disks (Giavalisco et al. 1996).
At redshifts lower than this, familiar 
Hubble types are in place with perhaps similar co-moving volume densities
at $z \sim 1$ as at $z \sim 0$ (van den Bergh et al. 2000). 
Clearly, identifying
the precursors to modern Hubble types at redshifts $1 < z < 2$ is critical
for understanding the physics behind galaxy formation.  By isolating and 
determining the properties of proto-ellipticals and proto-disks, we can begin 
to study physically the processes responsible for the formation of 98\% of the 
nearby bright galaxy population.  The problem is in identifying these systems.
Although clearly most galaxies at high redshift must somehow be the 
progenitors of modern galaxies, connecting specific high redshift populations
to low-redshift counterparts remains a complicated issue.

In this letter, we discuss the discovery and properties of a population of
luminous diffuse objects (LDOs) and luminous asymmetric objects (LAOs) 
at $z > 1$ in GOODS South HST imaging, and argue that these are
normal galaxies, disks and ellipticals, in the process of morphological 
formation. We identify LDOs and LAOs through their unique morphological
and structural properties, identified using the CAS morphological 
system (Conselice 2003), that differentiates them from modern
day Hubble types.  Similar galaxies appear to be missing in the same
abundance at $z > 2$ and at $z < 0.5$.  We discuss these 
objects, their physical properties, and briefly describe how they fit 
into the picture of how the Hubble sequence came into place.

\section{Data and Observations}

The data for this letter come from several sources, including Hubble
Space Telescope (HST) and
ground based imaging as part of the Great Observatories Origins Deep 
Survey (GOODS) (Giavalisco et al. 2003).  The major component of GOODS
used in this paper is moderately deep Hubble Space Telescope 
imaging of the Chandra Deep Field South region, covering $\sim$
150 arcmin$^{2}$.

The data set we use consists of the first three epochs of observations for
this field, resulting in images in the F435W (B), F606W (V), F775W (i)
and F850L (z) filters.  The separate frames within each band were combined
together, and mosaiced onto a large single image.  SExtractor was then
run on these mosaics to produce a catalog of galaxies, whose output 
includes structural information, such as ellipticity.  
The morphological analysis we perform is based on the model
independent CAS (concentration, asymmetry and clumpiness) 
parameters (Conselice 2003).   The SExtractor produced z-band catalog 
was used as input positions for the CAS program, and we used this same
positional catalog for each band.  

To obtain absolute magnitudes and redshifts, we use the photometric
redshifts computed by Mobasher et al. (2003) using the ACS images and 
ground based NIR and optical data from the GOODS ESO efforts. To minimize 
errors in the phot-zs, we only use galaxies to the
spectroscopic magnitude limit of the sample, R$_{\rm AB} \sim 25.5$, for which
we have an error estimate of $\delta(z)$/(1+$z_{\rm spec}$) = 0.11. This is 
the accuracy of the photometric redshifts, including
objects at $z > 1.5$, provided that they have R$_{\rm AB} <$ 25.5.

\section{Analysis} 

\subsection{Galaxy Identifications}

As argued in Conselice (2003), galaxies in different phases of evolution
can be uniquely identified through their structural properties (e.g.,
Conselice, Bershady \& Jangren 2000; Bershady, Jangren, \& Conselice 2000;
Conselice 2003).  In the CAS system early
type galaxies are those with high light concentrations ($C$),
low asymmetries ($A$) and low clumpiness values ($S$). Disk galaxies have lower
light concentrations, higher asymmetries and higher clumpiness values.  
Major mergers
can be identified through their large asymmetries (Conselice, Bershady
\& Gallagher 2000).

To understand the evolution of galaxy morphology, we ran the CAS morphological
code on all galaxies in the GOODS South
field detected with a signal to noise ratio of 5 or better, resulting in 
$\sim 10,000$ galaxies measured.   From
this list there are 2354 objects with M$_{\rm B} < -19$ that we use for
the analysis in this letter.  By choosing the somewhat luminous absolute
magnitude limit, we avoid incompleteness out to $z \sim 2$, based on
simulating nearby normal galaxies out to these redshifts (Conselice 2003;
Conselice et al. 2003).  We use BViz CAS values to interpolate
rest-frame B-band CAS parameters out to $z \sim 1.1$. At higher redshifts we
use the observed z-band values.  This introduces a possible morphological
bias which we address further in \S 3.3.

The concentration and
asymmetry parameters for all galaxies at $z < 2.5$ with M$_{\rm B} < -19$
are plotted in Figure~1.  There is a
distribution of points with averages $A = 0.26\pm0.10$ and 
$C = 2.69\pm0.45$.
This average concentration value happens to be very close to the value found
for an exponential profile, which is $C = 2.7$ (Bershady et al. 2000).
In this paper we study galaxies with concentration
values lower than 1 $\sigma$ from the average, or $C < 2.2$.  This
limit is somewhat arbitrary, although there are no nearby normal galaxies
with concentrations this low (Figure~1; Conselice 2003). 
We call these low concentrated galaxies luminous diffuse objects (LDOs)  
We separately examine galaxies with high asymmetries, $A > 0.35$,
with $A$ values larger than the clumpiness, or $A > S$.  
This asymmetry cut allows us to identify
galaxies undergoing tidal disturbances
(Conselice 2003).   We call these luminous asymmetric objects (LAOs). 
Images of LDOs and LAOs, as seen in a combined i+z mosaic, found at 
redshifts $z \sim 0.2$ to $z \sim 2$, are shown 
in Figure~2.  Note that simulations demonstrate
that galaxies do not become less concentrated or more asymmetric due to 
redshift effects out to $z \sim 3$ (Conselice 2003). That is, the fact
that we are seeing galaxies with low concentrations or high asymmetries
is not the result of lowered resolution or lowered S/N ratios. 

\subsection{Physical Properties of LDOs and LAOs: Star Formation}

Figure~3 shows the observe (i-z) color vs. redshift relationship for the LDOs
and LAOs with
Coleman, Wu and Weedman (1980) model SEDs for Sbc and Scd galaxy 
spectral types, and starburst templates from Kinney et al. (1996), 
over-plotted.  The LDOs and LAOs (i-z) colors match normal galaxies at 
z$ < 1$.  At 
higher redshifts these systems are clearly consistent with undergoing burst
of star formation.  In fact, these galaxies are a class of high redshift 
starbursts that are on average bluer
in (i-z) and (V-i) by $\sim 0.25$ mags than other bright non-LDO/LAO
galaxies at the same redshifts.

By using the approximate flux at rest-frame 1500 \AA\, emitted by these 
systems, through their observed B-magnitudes, we can calculate
their average star formation rates (Kennicutt 1998).  Doing this, we find
that the average star formation rates of the LDOs and LAOs, uncorrected for 
extinction, between $1.3<z<2$ are 4.0$\pm2.9$ and 3.6$\pm2.9$ \solm yr$^{-1}$,
compared
with 3.5$\pm2.4$ \solm yr$^{-1}$ for all galaxies at similar redshifts.  
While on average LDOs and LAOs do not have higher star formation
rates than other galaxies, the SF density in the LDOs and LAOs is 
roughly 35 - 40 \% of the total star formation rate between $1 < z < 2$.
These galaxies are therefore not only morphological peculiar, based on their
CAS parameters, but as a population they are undergoing large amounts of 
star formation, comparable to the amount found in bright star forming 
galaxies at $z \sim 2$ (Erb et al. 2003).   It is also interesting that
the LAOs have star formation rates that do not differ from 'average' 
galaxies in the $1.3<z<2$ redshift range.  This is possibly a sign that
the LAOs are pre-existing objects undergoing mergers with only modest
amounts of induced star formation (\S 3.4).
 
\subsection{Are LDOs 'Proto-Disks'}

How can this be answered, in the absence of kinematics?  One method 
is to determine if LDOs have properties similar to disks. 
Naturally, 
one of the ways to determine this is to look for spiral arms or bars in these 
galaxies. Figure~2a shows images of a small sample of the LDOs found at 
redshifts $z > 0.2$, where out to $z \sim 2.3$ spiral, and occasionally bar 
structures, can be seen.  
It is clear that the LDOs, especially those at $z < 1$ where
the signal to noise is higher, have bright, possibly
HII, regions in their outer parts.  These outer features are likely producing 
the low light concentration values.  The fact that there are relatively large 
numbers of LDOs that appear to be starbursting, with HII regions in their 
outer parts, and no obvious bulges in many cases, implies that some might
be disks forming through an outside-in process.

LDOs at $1 < z < 2$ also display a fairly strong absolute
magnitude-effective radius correlation (Figure~4), which is similar
in form to disks at $z < 1$ (Simard et al. 1999).  These
sizes are measured in the U-band, and may underestimate sizes 
that would be measured in the B-band, although starbursting galaxies look very 
similar between
these two wavelengths (Conselice et al. 2000c; Windhorst et al. 2002). Note
that most normal galaxies follow a magnitude-size relationship in the local
Universe, thus the fact that LDOs also follow this correlation
does not prove that they are disks. 
The straight line in Figure~4 shows the Freeman value between M$_{\rm B}$
and scale length, which disks at $z \sim 0$ scatter about.  From this, we 
can conclude that if any of the LDOs are disks, 
they must fade by $\sim 1.5$ magnitudes to fall on the Freeman relationship
at $z \sim 0$, assuming only passive luminosity evolution.

Another way to demonstrate that LDOs are likely not all small low-mass 
bursting galaxies is through their sizes. We can use their sizes to 
determine which $z \sim 0$ population LDOs might be 
evolving into, although size dispersions can be very large for
a given nearby morphological type
(Burstein et al. 1997).    LDOs are larger, on average, than 
typical Irr galaxies, and are most similar to nearby disks, although
the dispersion in $z \sim 0$ Irr sizes covers a large fraction of the observed
LDO sizes. The largest LDOs are not the 
largest galaxies between $1 < z < 2$, in fact they are roughly of average
size for galaxies with M$_{\rm B} < -19$.  LDOs are however larger than
low mass galaxies at $z \sim 0$.

There are roughly 1.8 LDOs arcmin$^{-2}$ in the GOODS South images.
LDOs however appear to be rare in the low-redshift and $z \sim 0$ 
Universe, and are most abundantly found between $z = 1 - 2$.  
The co-moving density of LDOs in this redshift range is 
$\sim$ 5 $\times 10^{5}$ Gpc$^{-3}$, similar to the local density of
bright disk galaxies. This number was calculated by finding the number of
LDOs at redshift intervals of equal co-moving volume 
(400 Mpc$^3$ arcmin$^{-2}$) out to $z \sim 2.5$. The co-moving density of LDOs 
decreases 
at lower redshifts, such that few are found at $z < 0.5$, although we cannot 
rule out that
volume effects or photometric redshift errors are producing this drop.   
Since we are examining galaxy morphologies
at $z > 1.1$ at rest wavelengths shorter than rest-frame B we could also
be introducing a morphological bias that makes galaxies appear less
concentrated and more asymmetric at short wavelengths.  To test 
whether the increase in the number of LDOs at higher redshift is the result 
of this
bandshifting, we redo the number density analysis using rest-frame 3000 \AA\
CAS values, and find the same qualitative behavior.  Thus morphological 
k-corrections are unlikely a source of significant bias.

\subsection{LAOs and LDOs as Proto-Spheroids}

Some LDOs and LAOs shown in Figure~2 could be identified as mergers, based
solely on their appearance.  We use the asymmetry index, which is suitable for
finding galaxies involved in major mergers (Conselice
2003; Conselice et al. 2003), to determine what fraction of
LDOs and LAOs could be involved in this process.  The average asymmetry values
of the LDOs are similar to the average asymmetry of all galaxies at the same 
redshifts.   Most of the highest
asymmetry LDOs are in fact at low redshift. Their high asymmetries
are likely due in part to the bright HII regions found in their outer 
parts.
This, their observed colors, and derived star formation rates reveal that 
these galaxies are involved in large
amounts of star formation which can produce high asymmetries (Conselice 2003).
While we cannot rule out that some LDOs are simply two or more galaxies
beginning to merge, the fact that they are less asymmetric than other galaxies
at the same redshift shows that they are not likely dominated by this process.

The LAOs are however likely mergers based on their
asymmetries (see Conselice et al. 2003).  By selecting these objects to have 
asymmetries
larger than their clumpiness vales, we are by design identifying galaxies
whose asymmetries are not produced solely by star formation, but also 
large scale tidal
effects (Conselice 2003).  The LAO images reveal systems that by eye
would be identified as mergers (e.g., LAO at $z = 0.53$), although some are 
spirals
that have been disturbed (e.g., LAO at $z =0.38$). Others appear to be 
lopsided early-types that
are possibly in the later stages of a merger (e.g., LAO at $z = 1.31$).  
All of these galaxy types
will be studied in detail in later papers.  The LAOs also do not follow the
same co-moving density evolution as the LDOs and their numbers peak at about
$z \sim 1$.   While it is premature to say at this point if these
galaxies are evolving into spheroids, it seems likely as many already
appear spheroidal, while others appear to be major mergers in progress.

The fact that LAOs have a peak density at $z \sim 1$ suggests that
some spheroidal formation may occur later than some disks, or even
from disks merging at $z > 1$.  Perhaps the LAOs identified here 
represent a secondary
episode of spheroid formation that produced younger field ellipticals.  
Future studies will
focus on the internal properties of these galaxies, including color structures,
and environment, to determine how these objects
fit into our picture of how the Hubble sequence formed.

Support for this work was provided by NASA through grant GO09583.01-96A 
from the Space Telescope Science Institute, which is operated by the 
Association of Universities for Research in Astronomy, under NASA contract 
NAS5-26555. 


\newpage

\begin{figure}
\plotfiddle{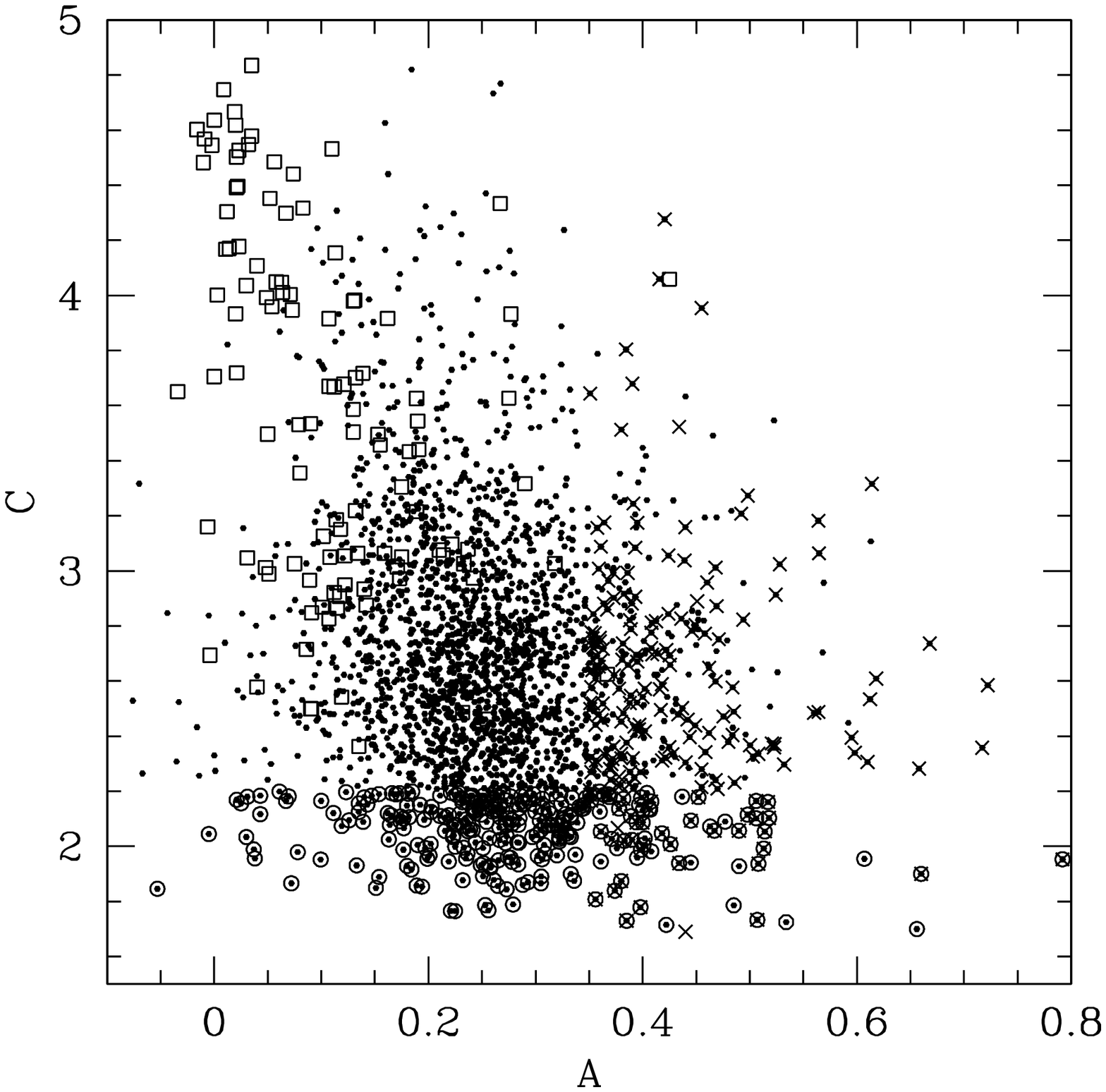}{6.0in}{0}{80}{80}{-250}{-100}
\vskip 0in
\caption{Rest-frame B-band C-A diagram where the selected LDOs are
circled, and the LAOs are crosses.  Values for nearby normal galaxies
are included as open boxes (Conselice 2003).}
\end{figure}

\begin{figure}
\plotfiddle{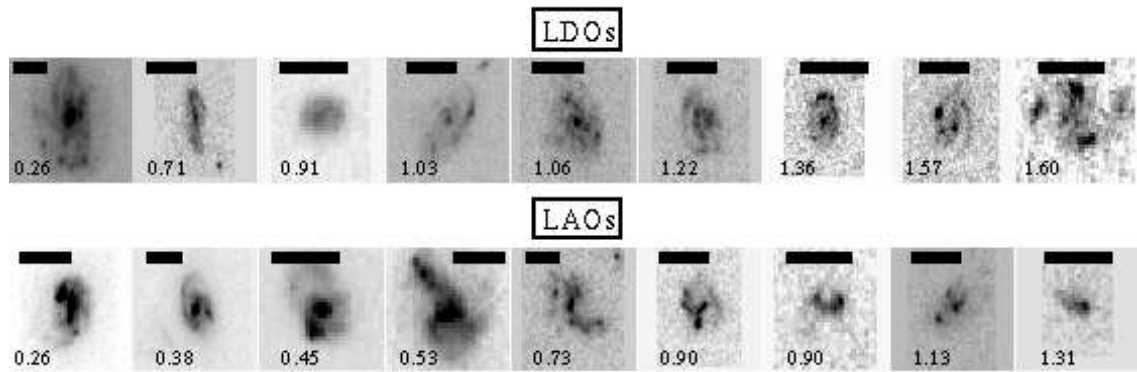}{6.0in}{0}{80}{80}{-250}{-100}
\vskip -1in
\caption{Images of (a) LDOs and (b) LAOs identified through their low 
light concentration index or high asymmetries in
the GOODS South ACS imaging.  The LDOs and LAOs are ordered by their 
redshift, labeled on each panel. The bar at the top of each image is
approximately 0.5\arcsec\, in length.}
\end{figure}

\begin{figure}
\plotfiddle{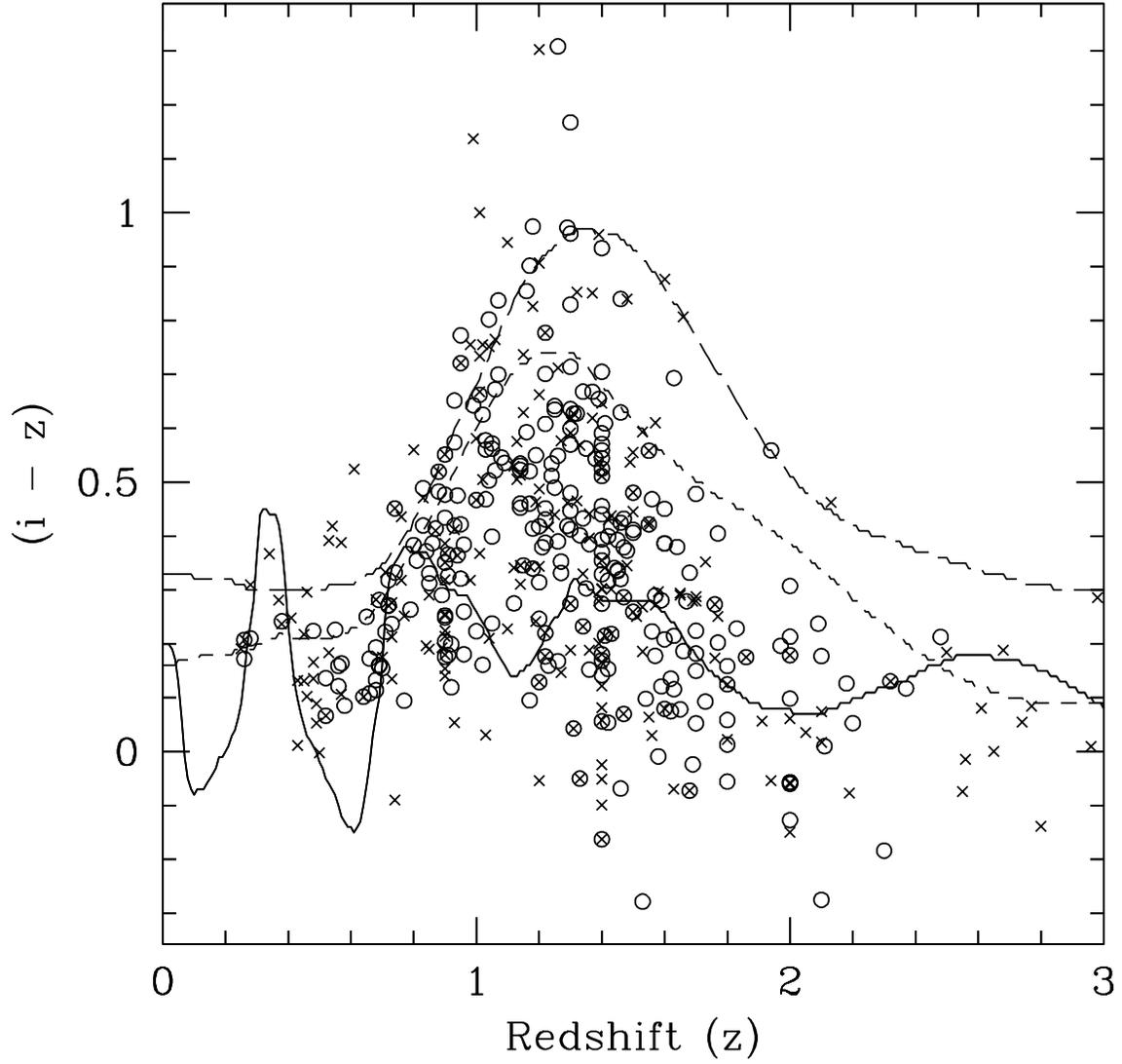}{6.0in}{0}{80}{80}{-250}{-100}
\vskip 0in
\caption{The distribution of (i-z) colors as a function of redshift with
two CWW spectral energy distributions and a Kinney et al. (1996) starburst
model plotted.  These are from bluest to
reddest - starburst (solid line), Scd (dashed), Sbc (long dashed). The LDOs 
are circles and the LAOs are crosses.  Other non-LDO and LAO objects are 
on average redder by $\sim 0.25$ magnitudes than the LDOs/LAOs.}
\end{figure}

\begin{figure}
\plotfiddle{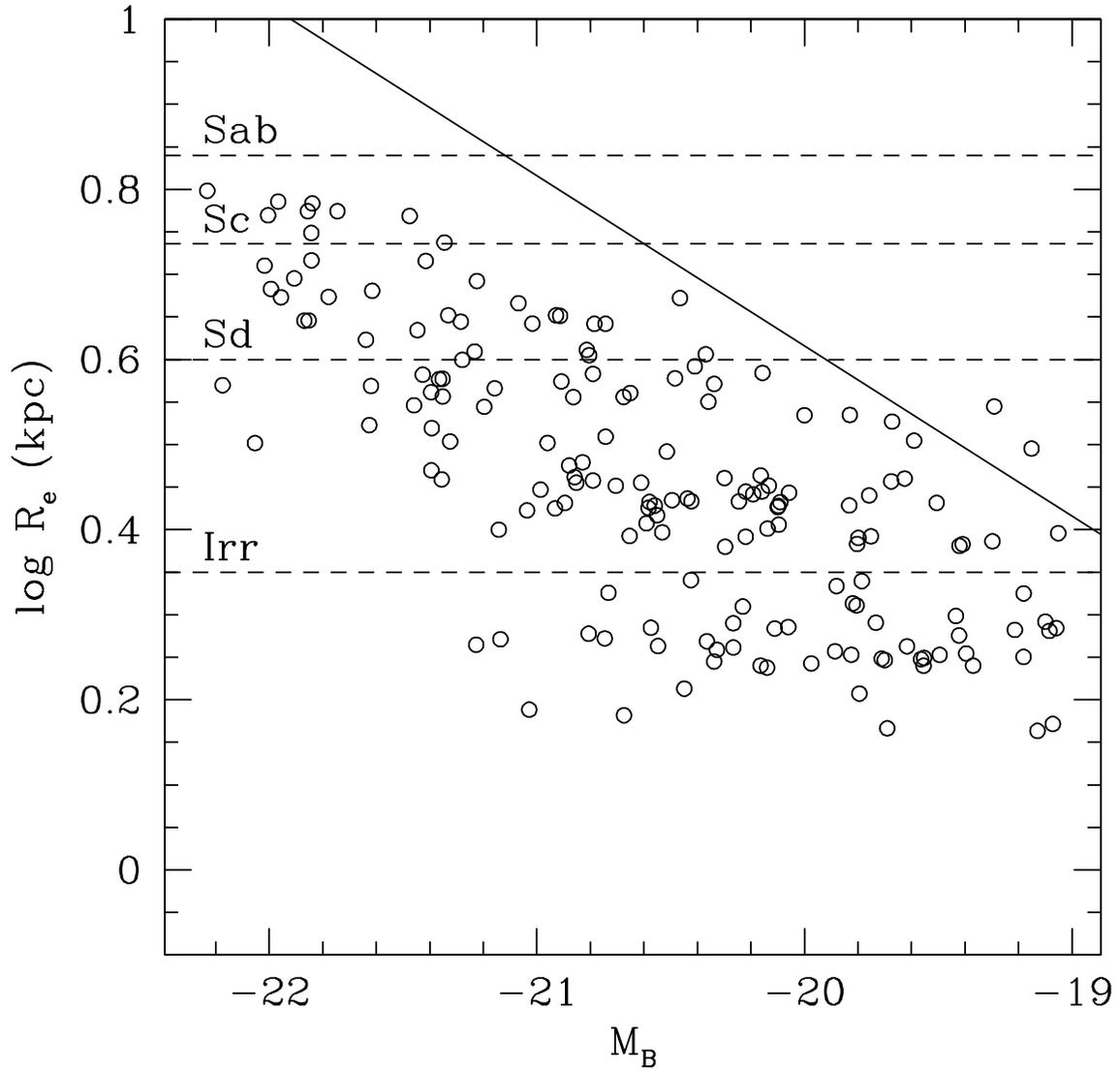}{6.0in}{0}{80}{80}{-250}{-100}
\vskip 0in
\caption{Absolute magnitude effective radius relationship for LDOs. The solid
line is the canonical Freeman disk relationship at $z \sim 0$.  The dashed
horizontal lines show the effective radii of different galaxy types
taken from Burstein et al. (1997).}
\end{figure}

\end{document}